\documentclass[%
 reprint,
 amsmath,amssymb,
 aps,
]{revtex4-1}

\usepackage{graphicx}
\usepackage{dcolumn}
\usepackage{bm}

\usepackage{xcolor}

\begin{document}

\preprint{APS/123-QED}

\title{Interaction-induced Interlayer Charge Transfer in the Extreme Quantum Limit}

\author{H.\ Deng, Y.\ Liu, I.\ Jo, L.N.\ Pfeiffer, K.W.\ West, K.W.\ Baldwin, and M.\ Shayegan}

\affiliation{Department of Electrical Engineering, Princeton University}

\date{\today}

\begin{abstract}

An interacting bilayer electron system
provides an extended platform to study electron-electron interaction beyond single layers.
We report here experiments demonstrating that the layer densities of an asymmetric bilayer electron system oscillate
as a function of perpendicular magnetic field that quantizes the energy levels.
At intermediate fields, this interlayer charge transfer can be well explained
by the alignment of the Landau levels in the two layers.
At the highest fields where both layers reach the extreme quantum limit,
however, there is an anomalous, enhanced charge transfer to the majority layer.
Surprisingly, when the minority layer becomes extremely dilute,
this charge transfer slows down as the electrons in the minority layer condense into a Wigner crystal.
Furthermore, by examining the quantum capacitance of the dilute layer at high fields,
the screening induced by the composite fermions in an adjacent layer is unveiled.
The results highlight the influence of strong interaction
in interlayer charge transfer in the regime of very high fields and low Landau level filling factors.

\end{abstract}

\pacs{Valid PACS appear here}
\maketitle

%

Low-disorder, interacting bilayer electron systems (BLESs) with their extra (layer) degree of freedom
provide a fascinating testbed for probing many-body physics.
Specifically, the charge distribution in a BLES with layers in close proximity
is directly influenced by the electron interaction.
For example, at zero magnetic field, one layer's density increases
when depleting the other by applying gate voltage
\cite{Eisenstein.PRB.50.1760, Katayama.SurfSci.305.405, Ying.PRB.52.R11611, Papadakis.PRB.55.9294, Footnote1}.
This phenomenon, observed in numerous BLESs
\cite{Eisenstein.PRB.50.1760, Katayama.SurfSci.305.405, Ying.PRB.52.R11611, Papadakis.PRB.55.9294, Footnote1, Larentis.NaLet.14.2039, Fallahazad.PRL.116.086601},
is known as negative compressibility (NC)
and reflects the dominance of the exchange and correlation energies as one layer becomes very dilute
\cite{Stern.PRB.30.840, Hedin.JPC.4.2064}.
Several studies have also reported interlayer charge transfer in the presence of a perpendicular magnetic field $B$
\cite{Eisenstein.PRB.50.1760, Davies.PRB.54.R17331, Manoharan.PRL.79.2722, Champagne.PRB.78.205310, Zhang.PRL.113.076804}.
Among these are experiments probing charge transfer at low magnetic fields \cite{Davies.PRB.54.R17331},
and also at high fields near particular Landau level (LL) filling factors,
where an interaction-induced spontaneous interlayer charge transfer,
which leads to the formation of fractional quantum Hall states with asymmetric charge distribution,
was observed \cite{Manoharan.PRL.79.2722}.
More recently, measurements near the filling factor one revealed an interlayer charge transfer
which was attributed to the formation of a Wigner crystal (WC) of quasi-particles \cite{Zhang.PRL.113.076804}.
There are also reports of charge transfer at very high magnetic fields
which we will discuss later in this paper in some detail \cite{Eisenstein.PRB.50.1760, Champagne.PRB.78.205310}.

Here we report a variety of interlayer charge transfers
in a large field range in an asymmetric BLES confined to a double quantum well (QW) [Fig. 1(a)].
At $B = 0$, the top-gate voltage ($V_{TG}$) dependence of densities shows the expected NC before the top layer is depleted.
At intermediate fields, the layers' densities oscillate as we increase $B$;
these can be described by a non-interacting model considering the alignment of the layers' LLs.
Most remarkable is the behavior at the highest $B$,
when both layers reach the extreme quantum limit (EQL)
so that only the lowest LL of each layer is occupied.
There is a surprisingly large charge transfer
from the top to bottom layer, beyond the prediction of LL alignment model.
Also anomalous is the dependence of $n_{T}$ on $V_{TG}$ in the EQL.
It first decreases faster than linearly as we lower $V_{TG}$, consistent with NC;
but in a wide range of $V_{TG}$ where the top layer becomes very dilute,
top layer remains finite density as the electrons condense into a WC.
An examination of the quantum capacitance of the top layer
points to the importance of screening by the bottom layer
which hosts composite fermions near filling factor 1/2.


%
\begin{figure*}[hbtp]
\includegraphics[width = 0.75\textwidth]{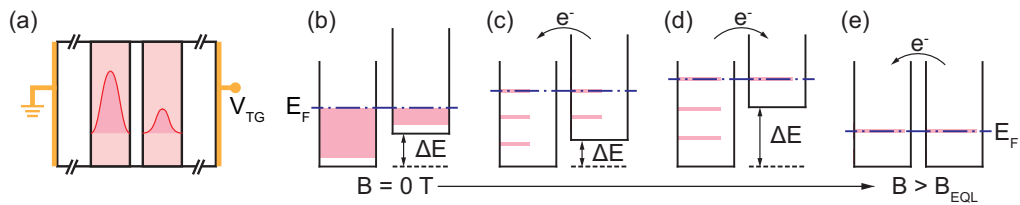}
\caption{
(a) Schematic of the sample structure.
The right and left pink-shaded regions indicate the top and bottom GaAs QWs
while the white regions represent the Al$_{0.24}$Ga$_{0.76}$As barriers.
Red curves show the charge distribution in the QWs.
$V_{TG}$ denotes the top-gate voltage.
(b)-(e) Landau level alignment and interlayer charge transfer induced by a perpendicular magnetic field ($B$).
$E_{F}$ is the Fermi level of the BLES,
and $\Delta E$ is the energy difference between the two layers.
$B_{EQL}$ is the field when both layers reach the EQL.
The arrows between the QWs indicate the direction of interlayer charge transfer compared to the case at $B = 0$ [Fig. 1(b)].
}
\end{figure*}
%

Our sample, grown by molecular beam epitaxy,
contains two 30-nm-wide GaAs QWs separated by a 10-nm-wide undoped Al$_{0.24}$Ga$_{0.76}$As barrier layer.
The QWs are modulation-doped with Si $\delta$-layers asymmetrically.
The top and bottom spacer layer thicknesses are 500 and 80 nm, respectively;
as grown,
the top-layer density
$n_{T} \simeq 0.3$ and
the bottom-layer density
$n_{B} \simeq 1.5$
in units of $10^{11}$ cm$^{-2}$
throughout the manuscript.
The sample has a van der Pauw geometry (4 mm $\times$ 4 mm)
with four In-Sn ohmic contacts on corners contacting both layers.
A deposited Ti-Au top gate and an In bottom gate are used to tune each layer's density.
In our experiments, the bottom gate is grounded
and only $V_{TG}$ is changed to tune $n_{T}$.
We determine $n_{T}$ and $n_{B}$ in the limit of zero-field, $n_{T,0}$ and $n_{B,0}$,
from the Fourier transform of Shubnikov-de Haas oscillations at very low $B$ ($\leq 0.5$ T).
At intermediate and high $B$, $n_{B}$ is determined
from the field positions of
the minima and maxima
in the measured longitudinal magnetoresistance ($R_{xx}$)
which primarily depends on $n_{B}$ because of the much lower $n_{T}$.
We use a low-frequency ($\leq$ 30 Hz) lock-in technique and a dilution refrigerator
with a base temperature of $\approx$ 30 mK.

We first present our observation and analysis
of the density oscillations with $B$.
To explain these oscillations,
we introduce a LL alignment (LLA) model.
At $B = 0$ [Fig. 1(b)], the Fermi levels ($E_{F}$) in two layers are aligned
as the system is at thermal equilibrium,
but there is an energy difference ($\Delta E$) between the conduction-band edges of the two layers
because of the unequal layer densities.
When $B$ is applied [Figs. 1(c - d)], the energy levels in the layers quantize into two sets of LLs.
Because of their different densities, each layer's $E_{F}$ may stay in a different LL,
but thermal equilibrium keeps
$E_{F}$
aligned \cite{Trott.PRB.39.10232, Solovyev.PRB.80.241310, Liu.PRL.107.266802}.
As a result, $\Delta E$ depends on $B$.
With increasing $B$, $\Delta E$ oscillates, causing a charge transfer
from the top to the bottom layer [Fig. 1(c)], or vise-versa [Fig. 1(d)].
For sufficiently high $B$,
when both layers enter the EQL [Fig. 1(e)],
$\Delta E$ equals zero and no longer changes, ending the charge transfer.

In a simple, classical picture, the amount of transferred charge after reaching the EQL is
$Q = (C\delta E)/e$, where $C = \epsilon/d$ is the interlayer capacitance,
$d$ is the interlayer distance,
$\epsilon$ is the dielectric constant,
and $\delta E$ is the difference between $\Delta E$ at a given $B$ and at $B = 0$.
More rigorously, the evolution of the LLs and interlayer charge transfer in a BLES is determined
by the subband densities rather than layer densities \cite{Davies.PRB.54.R17331, Liu.PRL.107.266802}.
In our sample, however, the subband densities are essentially the layer densities
because of the negligible interlayer tunneling and the strong asymmetry between the two layers \cite{Ying.PRB.52.R11611}.
The LLA model described above therefore provides a reasonably accurate description.
Indeed, in our experiments we find that
the simple LLA model semi-quantitatively explains the experimental data up to the EQL.


\begin{figure*}[htp]
\includegraphics[width=0.8\textwidth]{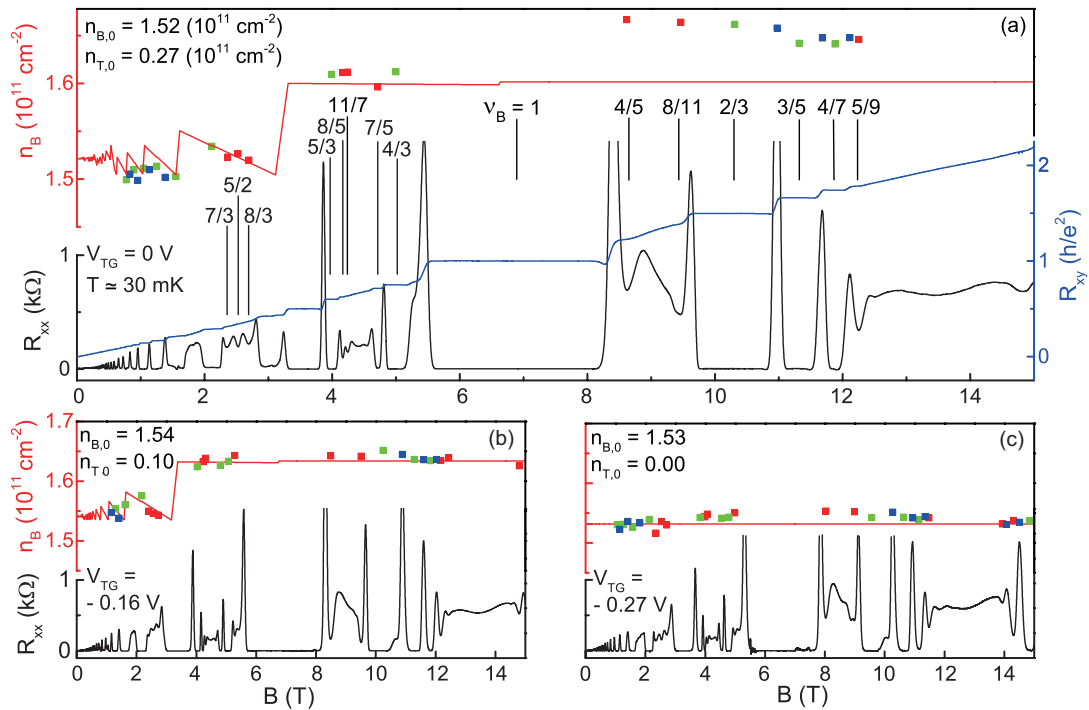}
\caption{
(a)-(c) Experimental data and calculations for different $n_{T}$ and $n_{B}$ at $B = 0$
($n_{T,0}$ and $n_{B,0}$, as indicated in each panel).
Black traces are the measured $R_{xx}$
and the blue trace [shown in (a)] is the Hall resistance $R_{xy}$.
Some typical QHSs observed in the bottom layer are marked with their filling factors ($\nu_{B}$) in (a).
Squares represent $n_{B}$ deduced from the following features in $R_{xx}$:
red squares are from the positions of QHSs with sharp minima
(e.g., $\nu_{B}$ = 5/2, 8/5, 5/9, etc.),
green squares are from the middle-point of a QHS's flat $R_{xx}$ minimum
(e.g., $\nu_{B}$ = 3, 2/3, etc.),
and blue squares are from the field position of the peak between adjacent QHSs
by assuming it represents an even-denominator filling of a LL
(e.g., $\nu_{B}$ = 9/2, 5/8, etc.).
Red lines represent $n_{B}$ expected from the LLA model.
Note that in panel (a) the experimentally deduced $n_{B}$ (squares) are
noticeably above the expected values when the EQL is reached at the highest $B$.
}
\end{figure*}


Our results are shown in Fig. 2
which presents $R_{xx}$ traces taken at different measured $n_{T,0}$ and $n_{B,0}$,
as listed in each of the panels.
The measured and calculated $n_{B}$ as $B$ increases are plotted
with square symbols and red lines
in the upper panels of Figs. 2(a - c).
We emphasize that the calculation only requires the sample structure information
and layer densities at $B = 0$ without fitting parameters.
The interlayer distance $d$ is defined as the QWs' center-to-center distance (40 nm),
and $\delta$-function-shaped LLs without broadening are used.


We start with describing Fig. 2(c)
where the top-layer is completely depleted at $B = 0$ ($n_{T,0} = 0$).
The sample should behave as a single-layer system.
As seen in Fig. 2(c),
both calculations and experimental data show that $n_{B}$ does not change in the full field range.
When the top layer is slightly populated at $B = 0$ [Fig. 2(b), $n_{T,0} = 0.10$],
calculations show that $n_{B}$ oscillates with $B$ because of the interlayer charge transfer.
As the LLA model predicts,
when $E_{F}$ lies in a LL, charge transfers
from the bottom to the top layer continuously with increasing $B$;
but once $E_{F}$ crosses the gap between LLs, charge transfers from the top to the bottom layer,
and the amount of transferred charge is proportional to the gap which $E_{F}$ crosses.
As Fig. 2(b) shows,
the experimental data quantitatively follow the calculations in the full range of $B$.

At yet higher $n_{T,0}$ [Fig. 2(a), $n_{T,0}$ = 0.27],
the calculations show a similar behavior for $n_{B}$ as in Fig. 2(b),
except that the amount of transferred charge is different.
In the low-$B$ regime (e.g., $\lesssim$ 3.5 T),
the experimental $n_{B}$ behaves as predicted by the calculations \cite{Footnote5_V3}.
However, when the bottom layer reaches the EQL ($B \gtrsim 7$ T),
the experimental $n_{B}$ is larger than the calculations predict,
evincing that the interlayer charge transfer is enhanced \cite{FootNote_g_factor}.


\begin{figure*}[ht]
\includegraphics[width = 0.8\textwidth]{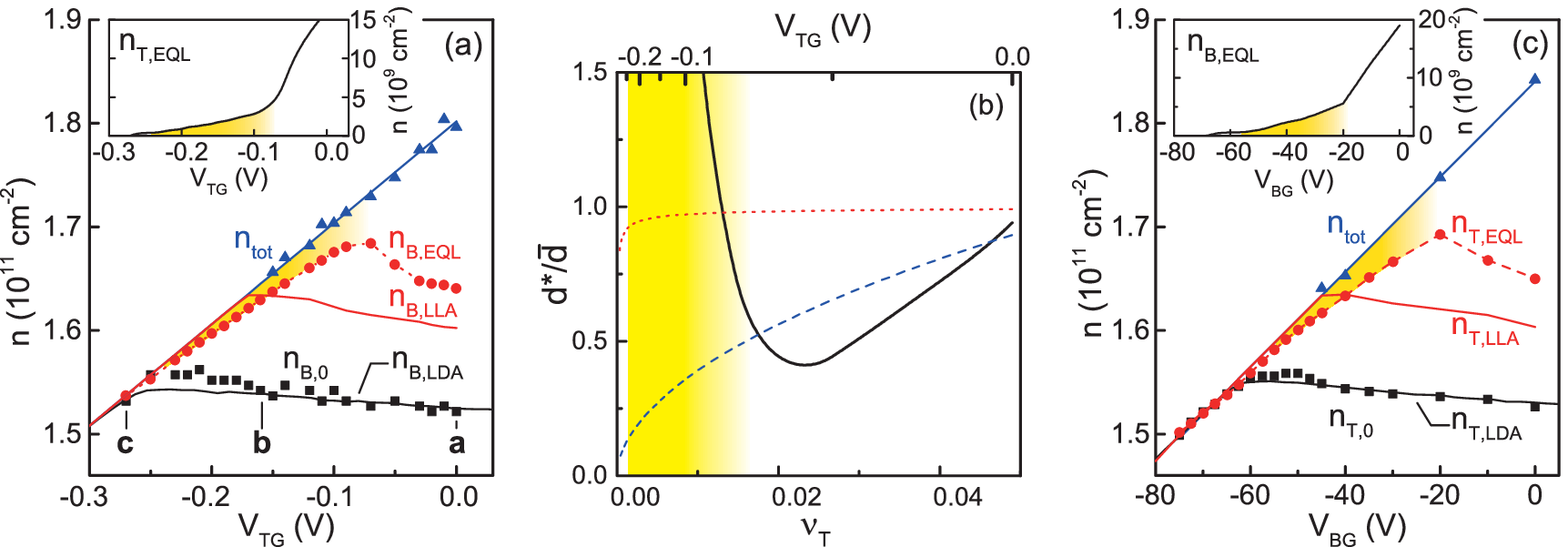}
\caption{
(a) Evolution of the densities with $V_{TG}$.
Black squares ($n_{B,0}$) are the bottom-layer densities at $B = 0$.
Black curve ($n_{B,LDA}$) represents $n_{B,0}$ from self-consistent, LDA calculations.
Red circles ($n_{B,EQL}$) are the measured $n_{B}$ in the EQL,
while the line ($n_{B,LLA}$) gives the expected $n_{B}$ in the EQL from the LLA model.
Blue triangles represent the measured $n_{tot}$,
and the blue line is a linear fit to the data points.
Note that $n_{tot} = n_{B,0} = n_{B,EQL}$ for $V_{TG} \leq -0.25$ V.
Markers a - c indicate $V_{TG}$ at which the data in Figs. 2(a - c) were taken.
Black curve in the inset ($n_{T,EQL}$) represents top-layer densities in the EQL.
(b) The normalized inverse capacitance $d^{*}/\bar{d}$ of the top layer from the experimental data (solid curve),
and theory with (blue-dashed curve) and without (red-dotted curve) screening effect
(see text for details).
(c) Evolution of the densities for another sample with a similar structure but inverted layer order.
All the definitions are analogous to those in Fig. 3(a).
In (a)-(c), the yellow-shaded areas indicate the regimes
where the Wigner crystal forms in the lower density layer.
}
\end{figure*}

To further highlight this enhanced interlayer charge transfer in the EQL systematically,
we summarize the $V_{TG}$ dependence of $n_{B}$ in Fig. 3(a).
The letters \textbf{a} to \textbf{c} in Fig. 3(a) mark $V_{TG}$ values at which data of Figs. 2(a - c) were taken.
The total density ($n_{tot}$) is the sum of $n_{T,0}$
[not plotted in Fig. 3(a)]
and $n_{B,0}$.
The average $n_{B}$ from the positions of the fractional QHS minima near $\nu_{B}$ = 1/2
gives $n_{B}$ in the EQL ($n_{B,EQL}$).
$n_{T}$ in the EQL ($n_{T,EQL}$) is deduced by subtracting $n_{B,EQL}$ from $n_{tot}$ [Fig. 3(a) inset].
The experimental $n_{tot}$ changes linearly with $V_{TG}$ as expected.
In contrast, $n_{B,0}$ \textit{increases} when we lower $V_{TG}$ down to $-0.25$ V.
This behavior is consistent with the NC of the system with a dilute top layer
\cite{Papadakis.PRB.55.9294, Ying.PRB.52.R11611}.
For further demonstration,
we also performed calculations,
solving Schr\"{o}dinger and Poisson equations self-consistently
in the local-density-approximation (LDA) \cite{Stern.PRB.30.840, Hedin.JPC.4.2064}.
The results [solid black curve marked $n_{B,LDA}$ in Fig. 3(a)]
reasonably agree with the measured $n_{B,0}$.

In Fig. 3(a), we also show a red curve marked $n_{B,LLA}$.
This curve is based on the LLA model
determining the expected $n_{B,EQL}$;
in Fig. 2 panels, $n_{B,LLA}$ are given by the red lines when the EQL is reached.
It is clear in Fig. 3(a) that the measured $n_{B,EQL}$ are higher than $n_{B,LLA}$
when the top layer is well populated ($V_{TG} > -0.16$ V),
thus highlighting the anomalous, enhanced charge transfer in the EQL.
This enhanced charge transfer suggests the important role of
electron-electron interaction which is not included in the LLA model.
As the NC phenomenon at $B = 0$ indicates
\cite{Eisenstein.PRB.50.1760, Katayama.SurfSci.305.405, Ying.PRB.52.R11611, Papadakis.PRB.55.9294,
Footnote1, Larentis.NaLet.14.2039, Fallahazad.PRL.116.086601, Stern.PRB.30.840, Hedin.JPC.4.2064},
this interaction has a significant effect on the charge distribution of a BLES.
A similar mechanism might be at work at high fields also \cite{Champagne.PRB.78.205310, Fano.PRB.37.8179, Footnote4}.
The exact form and the strength of the interaction at high fields,
however, is likely to be different from the $B = 0$ case,
and is presently unknown.

Even more intriguing is the dependence of $n_{B,EQL}$ on $V_{TG}$ seen in Fig. 3(a).
As $V_{TG}$ is decreased from its highest values,
$n_{B,EQL}$ initially $increases$ rapidly,
indicative of the very strong NC of the system in the EQL.
Note that the corresponding $n_{T,EQL}$ decrease faster than linearly with decreasing $V_{TG}$ [see Fig. 3(a) inset].
However, before the top layer is completely depleted and $n_{B,EQL}$ reaches $n_{tot}$,
the charge transfer slows down, and $n_{T,EQL}$ remains finite
in a relatively large range of $V_{TG}$ (down to $\sim -0.25$ V) [Fig. 3(a) inset].
This is not expected in a system with NC where the charge transfers in fact $accelerates$ just before the complete depletion
[e.g., $n_{B,0}$ in Fig. 3(a)].

We suggest that the retention of charge by the dilute top layer,
marked by the yellow-shaded regime in Fig. 3(a),
is linked to the formation of a magnetic-field-induced WC in this layer.
Our reasoning is based on Fig. 3(c)
which shows nearly identical data for a sample with a similar structure but an inverted layer order,
i.e. high $n_{T}$ and low $n_{B}$.
As described in Ref. \cite{Hao.PRL.117.096601},
the sample of Fig. 3(c) exhibits clear signatures of WC in the low-density layer (in this case, the bottom layer)
in the range of $-55\leq V_{BG}\leq -40$ V,
inside the yellow-shaded regime of Fig. 3(c).
These signatures are weak $R_{xx}$ maxima observed near the half-filling of the high-density (top) layer,
and can be attributed to the commensurability oscillations of the composite fermions (CFs) in the top layer,
induced by the periodic potential of the WC formed in the bottom layer \cite{Hao.PRL.117.096601}.
Unfortunately, we do not observe such commensurability oscillations in the sample of Fig. 3(a),
likely because of its large (4 mm $\times$ 4 mm) size and possible inhomogeneity;
as discussed in Ref. \cite{Hao.PRL.117.096601},
clear commensurability features are only observed
in small Hall bar samples with $< 1$ mm dimensions \cite{Footnote6}.

To further illustrate the behavior of the system in the EQL and also compare it to previous reports
\cite{Eisenstein.PRB.50.1760, Skinner.PRB.87.035409},
we derived our sample's differential capacitance
$C^{*} = \epsilon/d^{*} = (\partial n_{T,EQL}/\partial V_{TG})e$
from the experimental data of Fig. 3(a) inset.
We then normalized $C^{*}$ to the standard parallel-plate capacitance
between the top gate and the top layer in our sample,
and present $d^{*}/\bar{d}$ vs $\nu_{T}$ as a black curve in Fig. 3(b).
Here $\bar{d}$ is the distance between the top gate and the the top layer,
$\nu_{T}$ is the top-layer filling factor when $\nu_{B} = 1/2$,
and $d^{*}/\bar{d} < 1$ corresponds to NC \cite{Eisenstein.PRB.50.1760}.
Note the non-monotonic behavior of the experimental $d^{*}/\bar{d}$ vs $\nu_{T}$ curve.
The decrease of $d^{*}/\bar{d}$ for $\nu_{T} \gtrsim 0.02$
can be understood as the enhanced NC at high fields \cite{Eisenstein.PRB.50.1760}.
Indeed, in Ref. \cite{Eisenstein.PRB.50.1760}, reasonably good agreement is found
between the experimental data and the calculations for a single, dilute, layer \cite{Fano.PRB.37.8179}.
Applying this model to our system, however, leads to the red-dotted curve in Fig. 3(b)
which deviates significantly from our data.
The discrepancy mainly stems from ignoring the role of screening,
which is crucial in our BLES where the dilute layer is in close proximity to a layer of compressible CFs at $\nu_{B}=1/2$.
We therefore apply an approximate model from Ref. \cite{Skinner.PRB.87.035409},
modified for our sample geometry \cite{Footnote3_V2}.
The model describes the dependence of the quantum capacitance in the EQL of a dilute layer,
which is screened by an adjacent perfect metal layer.
In Fig. 3(b), we present the prediction of this model (blue-dashed curve),
showing much better agreement with the experimental data for $\nu_{T} \gtrsim 0.02$.

When $\nu_{T} \lesssim 0.02$, our experimental data show a rise in $d^{*}/\bar{d}$.
This is qualitatively opposite to the theoretical prediction that
$d^{*}/\bar{d}$ should monotonically decrease and approach zero as $\nu_{T}$ goes to zero
even as the 2D electrons crystallize into a WC \cite{Skinner.PRB.87.035409}.
The observation of a rise in $d^{*}/\bar{d}$ at very low filling factors
was also reported in compressibility measurements \cite{Eisenstein.PRB.50.1760},
and has been attributed to the density fluctuations caused by disorder
\cite{Eisenstein.PRB.50.1760, Skinner.PRB.87.035409}.
In our samples, however, in the same low $\nu_{T}$ parameter range where $d^{*}/\bar{d}$ is rising,
e.g., for $\nu_{T} \simeq 0.015$, we observe clear signs of WC order \cite{Hao.PRL.117.096601, Footnote6}.
This observation strongly suggests that the increase of $d^{*}/\bar{d}$ at low fillings
does not necessarily correspond to the random localization of electrons in a disorder potential.

We acknowledge the NSF (Grant DMR 1305691) for measurements,
and the Gordon and Betty Moore Foundation (Grant GBMF4420),
the DOE BES (Grant DE-FG02-00-ER45841),
and the NSF (Grants MRSEC DMR 1420541 and ECCS 1508925) for sample fabrication.

\bibliographystyle{h-physrev}

\begin{thebibliography}{10}
\bibitem{Footnote1}
A qualitatively similar charge transfer is also seen at high magnetic fields
  \cite{Eisenstein.PRB.50.1760}; we will discuss this in detail later in the
  paper.
\bibitem{Footnote5_V3}
When the top layer is well populated [Fig. 2(a)], we observe an anomalous
  feature (maximum) in the $\nu_{B}=2$ plateau of $R_{xx}$ (near $B=3.3$ T).
  This feature reflects the presence of the finite-density top layer, and does
  not affect our measurement of $n_{B}$ at higher fields where we do not see
  anomalous features.
\bibitem{FootNote_g_factor}
Assuming an enhanced effective $g$-factor would lead to a larger energy gap and
  a more significant charge transfer at $\nu_{B} = 1$ in the LLA calculations.
  However, as described in the text, the total amount of transferred charge
  after reaching the EQL relative to $B = 0$ is determined by the difference
  between $\Delta E$ in the EQL and at $B=0$, and thus $n_{B}$ in the EQL would
  not increase with an enhanced $g$-factor. Since the value of the $g$-factor
  at $B = 0$ and its dependence on $B$ are not quantitatively known, we used
  the band-value of GaAs $g$-factor ($-0.44$).
\bibitem{Footnote4}
In Ref. \cite{Champagne.PRB.78.205310}, an estimate based on the "backbone"
  energy dependence on filling factor \cite{Fano.PRB.37.8179} was applied to
  explain the enhanced imbalance at total filling factor one in a BLES. The
  results highlight the importance of the electron-electron interaction at high
  fields.
\bibitem{Footnote3_V2}
The theory of Ref. \cite{Skinner.PRB.87.035409} considers the capacitance
  between a layer of WC at high $B$ and a perfect metal layer at a close
  distance $\bar{d}$ which screens the electron interaction in the WC layer. An
  analytical expression [Eq. (9) in \cite{Skinner.PRB.87.035409}] can be
  deduced from this model in the limit of $\bar{d}/l_{B} \ll 1$ or $\nu_{WC}
  \ll l_{B}^{2}/\bar{d}^{2}$, where $l_{B}$ is magnetic length and $\nu_{WC}$
  is the filling factor of the WC layer. This analytical expression is not
  directly applicable to our samples where $\bar{d}$ ($\simeq 750$ nm) is about
  100 times larger than $l_{B} (\simeq 7$ nm). However, in our system the
  bottom layer which hosts CFs near $\nu_{B} = 1/2$ is close and can provide
  screening. To calculate the dashed curve shown in Fig. 3(b), we used $\bar{d}
  = 25$ nm in Eq. (9) of Ref. \cite{Skinner.PRB.87.035409} as the effective
  interlayer distance; this $\bar{d}$ is somewhat smaller than our QWs'
  center-to-center distance ($d = 40$ nm) to account for the finite thickness
  of layer charge distributions in our QWs. Note that with this $\bar{d}$, the
  condition $\nu_{WC} \ll l_{B}^{2}/\bar{d}^{2}$ is satisfied in our
  experimental range. Our purpose of showing the dashed curve is mainly to
  emphasize the qualitative contrast between the experimental data and the
  prediction of the calculations.
\end{thebibliography}

\end{document}